\documentstyle[12pt]{article}
\begin{document}
\hfill{CCUTH-97-06}
\vfill
\centerline{\large\bf Perturbative pion form factor reexamined}
\vskip 1.5cm
\centerline{David Tung and Hsiang-nan Li}
\vskip 0.5cm
\centerline{Department of Physics, National Chung-Cheng University,}\par
\centerline{Chia-Yi, Taiwan, Republic of China}\par
\vskip 1.0cm

\centerline{\bf Abstract}
\vskip 1.0cm

We recalculate the pion electromagnetic form factor based on the
perturbative QCD formalism that includes the Sudakov resummation. We take
into account the evolution of the pion wave function in $b$, which
represents the transverse extent of the pion. An infrared enhancement is
observed when $1/b$ approaches $\Lambda_{\rm QCD}$. We propose to freeze
the evolution of the wave function at some scale above $\Lambda_{\rm QCD}$,
such that perturbative contributions are stablized. Our predictions are
consistent with experimental data, and insensitive to the variation of
relevant parameters.

\vfill

\newpage

\centerline{\large \bf I. Introduction}
\vskip 0.5cm

The applicability of perturbative QCD (PQCD) to exclusive processes has been
a controversy for almost two decades \cite{IL1,HS}. Although there is a
general agreement that PQCD can successfully make predictions for exclusive
reactions as momentum transfers go to infinity, it remains unclear
whether experimentally accessible energy scales are large enough to justify
these predictions. A progress was made recently by Li and Sterman
\cite{LS}, who proposed a modified PQCD formalism with Sudakov suppression
taken into account. They found that nonperturbative contributions are
suppressed significantly by the Sudakov effects, and the applicability of
PQCD can be extended down to few GeV scales. However, their analysis of the
pion electromagnetic form factor did not concern the match with experimental
data, such that a convincing piece of justification of their approach is
missing. In this paper we shall further include the evolution of the pion
wave function into the modified PQCD formalism, and demonstrate that our
predictions explain the data.

The PQCD theory for exclusive processes was first proposed by Brodsky and
Lepage \cite{LB1}. They argued that an exclusive process, such as the pion
form factor, can be factorized into two types of
subprocesses: wave functions which carry the nonperturbative information of
the initial and final state pions, and a hard amplitude which
describes the scattering of a valence quark of the pion off the energetic
photon. The former is not calculable in perturbation theory, and needs to be
parametrized by a model, to be derived by nonperturbative methods, such as
QCD sum rules and lattice gauge theory, or to be determined by experimental
data. The latter, characterized by a large momentum flow, is calculable in
perturbation theory. Combining these two types of subprocesses, predictions
for the pion form factor at experimentally accessible energies scales can 
be made.

According to this naive picture, the factorization formula for the pion
form factor $F_\pi(Q^2)$, graphically represented by Fig.~1(a), is written
as
\begin{equation}
F_\pi(Q^2)=\int_0^1 d x_1d x_2\phi(x_2,\mu)H(x_1,x_2,Q,\mu)\phi(x_1,\mu) \;.
\label{pi}
\end{equation}
$Q^2=-2P_1\cdot P_2$ is the momentum transfer from the photon, $P_1$ ($P_2$)
being the mometum of the incoming (outgoing) pion. $\mu$ is a renormalization
and factorization scale, below which QCD dynamics is regarded as being
nonperturbative and absorbed into the pion wave function $\phi$, and above
which QCD dynamics is regarded as being perturbative and absorbed into the
hard amplitude $H$. $\phi(x,\mu)$ gives the probability of a valence quark
carrying a fractional momentum $xP$ in the parton model at the scale $\mu$.
Usually, the $\mu$ dependence of $\phi$ is neglected, {\it i.e.},
$\phi(x,\mu)=\phi(x)$. $H$ is obtained by computing quark-photon scattering
diagrams.

To make predictions for $F_\pi$, one substitutes the lowest-order expression
of $H$ derived from Fig.~1(b),
\begin{equation}
H(x_1,x_2,Q,\mu)=\frac{16\pi C_F \alpha_s(\mu)}{x_1x_2Q^2}\;,
\label{pih}
\end{equation}
and the asymptotic pion wave function \cite{LB2}
\begin{equation}
\phi^{AS}(x)=\frac{3f_{\pi}}{\sqrt{2N_{c}}}x(1-x)\;,
\label{as}
\end{equation}
into Eq.~(\ref{pi}), where $\alpha_s(\mu)$ is the running coupling constant,
$C_F=4/3$ the color factor, $N_{c}=3$ the number of colors, and
$f_{\pi}=0.133$ GeV the pion decay constant. Because $\phi^{AS}$ peaks
at $x=1/2$, the main contributions to Eq.~(\ref{pi}) come from
the region with intermediate $x$. Higher-order corrections to
$H$ then produce logarithms of the type $\ln(Q^2/\mu^2)$, which may be
so large as to spoil the perturbative expansion. To eliminate the
logarithms, a natural choice of $\mu$ is $\mu=Q$. It is easy to find that
$Q^2F_\pi(Q^2)\sim 0.12$ GeV$^2$ for $Q^2\sim 4$-10 GeV$^2$ is only
1/3 of the data $\sim 0.35$ GeV$^2$ \cite{d1,d2,d3,CJB}. This contradiction
implies that the pion form factor has not yet become completely asymptotic
at experimentally accessible energies.

Hence, one may resort to a preasymptotic model, the Chernyak and Zhitnitsky
(CZ) wave function derived from QCD sum rules \cite{CZ1},
\begin{equation}
\phi^{CZ}(x)=\frac{15f_{\pi}}{\sqrt{2N_{c}}}x(1-x)(1-2x)^{2}\;,
\label{cz}
\end{equation}
which possesses maxima at the end points $x\to 0$ and $x\to 1$. 
The CZ wave function enhances the predictions for $Q^2F_\pi(Q^2)$ to
0.30 GeV$^2$, and improves the match with the data. Howerer, such a success
was criticized by Isgur and Llewellyn Smith \cite{IL1,IL2} and by Radyushkin
\cite{BR}: Since $\phi^{CZ}$ emphasizes the contributions from small $x$,
higher-order corrections to $H$ in fact produce logarithms like
$\ln(x_1x_2Q^2/\mu^2)$. Hence, the natural
choice of $\mu$ is $\mu=\sqrt{x_1x_2}Q$, which is consistent with
$\mu=Q$ employed above for intermediate $x$, but results in a large coupling
constant $\alpha_s(\sqrt{x_1x_2}Q)>1$ at small $x$, indicating that the
end-point regions are nonperturbative regions. A careful analysis revealed
that the enhancement of the predictions for $Q^2F_\pi(Q^2)$
is due to the amplification of the nonperturbatve end-point contributions.
As a consequence, the perturbative calculation loses its self-consistency as
a weak-coupling expansion.

The above discussions hint that the end-point regions should be treated in a
different way. In general, a valence quark in the pion can carry a small
amount of transverse momenta $k_T$, which then flow through the hard
scattering. Therefore, the hard gluon propagator is written as 
\begin{equation}
\frac{1}{x_1x_2Q^2+k_T^2}\approx
\frac{1}{x_1x_2Q^2}\left(1-\frac{k_T^2}{x_1x_2Q^2}\right)\;,
\end{equation}
where the first term corresponds exactly to the asymptotic $H$ in 
Eq.~(\ref{pih}), and the second is suppressed by powers of $Q^{2}$ at 
intermediate $x$. If the
end-point regions are not important, one may drop the second term, and
integrate out the $k_T$ dependence in the wave functions, arriving at
the leading-power factorization formula in Eq.~(\ref{pi}). However, the
end-point difficulities indicate that the higher-power effects are crucial,
and should be kept at the outset as deriving the perturbative expression for
the pion form factor.

The introduction of $k_T$ leads to three observations
immediately. First, the pion form factor becomes a two-scale ($Q$ and $k_T$)
problem, and the resummation technique is required to organize the large
logarithms $\ln(Q^2/k_T^2)$ from radiative corrections to the wave
functions. Second, the process must be analyzed in the Fourier transform
space of $k_T$, denoted by $b$ \cite{CS}, which is regarded as the
transverse separation between the valence quarks of the pion. Third, the
quantity $1/b$, now considered as one of the characteristic scales of the
hard amplitude, should be substituted for the argument of
$\alpha_s$ if $1/b > \sqrt{x_1x_2}Q$. 

Li and Sterman found that the resummation of the large logarithms suppresses 
the elastic scattering at large spatial separation. This property,
called Sudakov suppression \cite{CS,BS,JCC}, makes nonperturbative
contributions from large $b$, no matter what $x$ are, less important, and
improves the applicability of PQCD at few GeV scales. However, their
discussion was restricted to the PQCD applicability, such that the
results of the pion form factor do not match the data. In the present
work we shall aim to explain the experimental data by including a new
piece of information, the evolution of the wave function, into the modified
PQCD formalism. Our predictions are found to be dominated by perturbative
contributions for $Q^2 > 4$ GeV$^2$, and in good agreement with the data.

A brief review of the standard asymptotic
expression for the form factors \cite{LB2,CZ1,FJ,ER,CZ2,CG} may help to
motivate our viewpoint. The factorization formula for the pion form factor
including the evolution of the wave function is derived in Sect. II.
Numerical results are given in Sect. III. Section IV is the conclusion.

\vskip 1.0cm

\centerline{\large \bf II. The Factorization Formula}
\vskip 0.5cm

In this section we review factorization theorems including Sudakov effects
for the pion form factor, which is expressed as the convolution of a hard
amplitude with wave functions. We investigate radiative corrections to the
quark-photon scattering diagrams shown in Fig.~2, and explain how they are 
factorized
into the convolution factors. There are two types of important corrections:
collinear, when the loop momentum is parallel to the incoming or outgoing
pion momentum, and soft, when the loop momentum is much smaller than the
momentum transfer $Q^2$. Each type of the important corrections gives rise
to large single logarithms. They may overlap to produce double logarithms in
some cases. In axial gauge the two-particle reducible diagrams, like
Figs.~2(a) and 2(b), give double logarithms from the overlap of collinear and
soft divergences, while the two-particle irreducible corrections, like
Figs.~2(c) and 2(d), give only single soft logarithms.

It can be shown that soft divergences cancel between Figs.~2(a) and 2(b), as
well as between 2(c) and 2(d), in the asymptotic region with $b\to 0$.
Therefore, reducible corrections are dominated by collinear divergences, and
can be absorbed into the pion wave funtion $\cal P$, which involves similar
dynamics. Irreducible corrections, due to the cancellation of soft
divergences, are then absorbed into the hard scattering amplitude $H$.
Based on the above reasoning, the factorization formula for the pion form
factor is written in the $b$ space as \cite{LS},
\begin{eqnarray}
F_{\pi}(Q^{2})&=&\int_0^1 d x_{1}d x_{2}\int
\frac{d^2 {\bf b}}{(2\pi)^2}
{\cal P}(x_{2},b,P_{2},\mu)
\nonumber \\
& &\times \,{\tilde H}(x_1,x_2,b,Q,\mu)
{\cal P}(x_{1},b,P_{1},\mu)\; ,
\label{pi1}
\end{eqnarray}
where $\tilde H$ is the Fourier transform of $H$. 

We choose the Breit frame such that $P_1^+=P_2^-=Q/\sqrt{2}$ and all other
components of $P$'s vanish. After resumming the double logarithms, we
obtain
\begin{equation}
{\cal P}(x,b,P,\mu)=\exp\left[-s(x,b,Q)-s(1-x,b,Q)\right]
{\bar{\cal P}}(x,b,\mu)\; .
\label{sp}
\end{equation}
The exponent $s(\xi,b,Q)$, $\xi=x$ and $1-x$, is expressed as \cite{BS}
\begin{equation}
s(\xi,b,Q)=\int_{1/b}^{\xi Q/\sqrt{2}}\frac{d p}{p}
\left[\ln\left(\frac{\xi Q}
{\sqrt{2}p}\right)A(\alpha_s(p))+B(\alpha_s(p))\right]\;,
\label{fsl}
\end{equation}
where the anomalous dimensions $A$ to two loops and $B$ to one loop are
given by
\begin{eqnarray}
A&=&{\cal C}_F\frac{\alpha_s}{\pi}+\left[\frac{67}{9}-\frac{\pi^2}{3}
-\frac{10}{27}n_f+\frac{8}{3}\beta_1\ln\left(\frac{e^{\gamma_E}}{2}\right)
\right]\left(\frac{\alpha_s}{\pi}\right)^2\;,
\nonumber \\
B&=&\frac{2}{3}\frac{\alpha_s}{\pi}\ln\left(\frac{e^{2\gamma_E-1}}
{2}\right)\;,
\end{eqnarray}
$\gamma_E$ being the Euler constant. The two-loop running coupling constant,
\begin{equation}
\frac{\alpha_s(\mu)}{\pi}=\frac{4}{\beta_0\ln(\mu^2/\Lambda^2)}-
\frac{16\beta_1}{\beta_0^3}\frac{\ln\ln(\mu^2/\Lambda^2)}
{\ln^2(\mu^2/\Lambda^2)}\;,
\label{ral}
\end{equation}
with the coefficients
\begin{eqnarray}
& &\beta_{0}=\frac{33-2n_{f}}{3}\;,\;\;\;\beta_{1}=\frac{153-19n_{f}}{6}\;,
\label{12}
\end{eqnarray}
and the QCD scale $\Lambda\equiv \Lambda_{\rm QCD}$, will be substituted
into Eq.~(\ref{fsl}). We require the relation of the involved scales
$\xi Q/\sqrt{2} > 1/b > \Lambda$ as indicated by the bounds of the variable
$p$ in Eq.~(\ref{fsl}). QCD dynamics below $1/b$ is regarded as being
nonperturbative, and absorbed into the initial condition
${\bar{\cal P}}(x,b,\mu)$.

The functions ${\bar{\cal P}}$ and $\tilde H$ still contain single logarithms
from ultraviolet divergences, which can be summed using the RG methods
\cite{BS}. The large-$b$ behavior of $\cal P$ is then summarized as
\begin{eqnarray}
{\cal P}(x,b,P,\mu)&=&\exp\left[-s(x,b,Q)-s(1-x,b,Q)
-2\int_{1/b}^{\mu} \frac{d\bar{\mu}}{\bar{\mu}}\gamma
_q(\alpha_s(\bar{\mu}))\right]
\nonumber \\
& &\times {\bar{\cal P}}(x,b,1/b)\;,
\label{pb}
\end{eqnarray}
with $\gamma_q=-\alpha_s/\pi$ the quark anomalous dimension in axial
gauge. To concentrate on the evolution of ${\bar{\cal P}}$ in $1/b$, we
ignore the intrinsic $b$ dependence denoted by the argument $b$ \cite{BSW},
that is, we adopt ${\bar{\cal P}}(x,b,1/b)=\phi(x,1/b)$.

Since the pion form factor, as a physical observable, is $\mu$ independent,
the RG analysis applied to $\tilde H$ gives
\begin{eqnarray}
{\tilde H}(x_i,b,Q,\mu)
&=&\exp\left[-4\,\int_{\mu}^{t}\frac{d\bar{\mu}}{\bar{\mu}}
\gamma_q(\alpha_s(\bar{\mu}))\right]\nonumber \\
& &\times {\tilde H}(x_i,b,Q,t)\;,
\label{13}
\end{eqnarray}
where $t$ is the largest mass scale involved in the hard scattering,
\begin{equation}
t=\max(\sqrt{x_{1}x_{2}}Q,1/b)\; .
\label{9}
\end{equation}
The scale $\sqrt{x_1x_2}Q$ is associated with the longitudinal momentum of
the hard gluon and $1/b$ with the transverse momentum. At lowest order, $H$
is given by
\begin{eqnarray}
H(x_i,{\bf k}_{T_i},Q) &=&
\frac{16\pi\alpha_s{\cal C}_Fx_{1}Q^{2}}
{(x_{1}Q^{2}+{\bf k}_{1T}^{2}) (x_{1}x_{2}Q^{2}+
({\bf k}_{1T}-{\bf k}_{2T})^2)}
\label{hna} \\
&\sim &\frac{16\pi\alpha_s{\cal C}_F}
{x_{1}x_{2}Q^{2}+({\bf k}_{1T}-{\bf k}_{2T})^{2}}\; ,
\label{7}
\end{eqnarray}
where we have neglected the transverse momentum in the numerator. In the 
second form, we have further neglected the transverse momentum associated 
with the virtual fermion lines in $H$, which are linear rather than 
quadratic in $x$. Because Eq.~(\ref{7}) depends on the combination of 
transverse momenta, $\tilde H$ involves only a single $b$, which has been 
made explicit in Eq.~(\ref{pi1}).

Combining all the above ingredients, we arrive at the factorization formula
for the pion form factor,
\begin{eqnarray}
F_{\pi}(Q^2)&=& 16\pi{\cal C}_F\int_{0}^{1}d x_{1}d x_{2}
\int_{0}^{\infty} bd b\phi(x_{1},1/b)\phi(x_{2},1/b)
\alpha_{s}(t) K_{0}(\sqrt{x_{1}x_{2}}Qb)
\nonumber \\
& &\times \exp[-S(x_{1},x_{2},b,Q)]\;.
\label{15}
\end{eqnarray}
where the complete Sudakov exponent is given by
\begin{equation}
S(x_{1},x_{2},b,Q)=\sum_{i=1}^{2}\left[s(x_{i},b,Q)+s(1-x_{i},b,Q)\right]+
4\int_{1/b}^{t}\frac{d\bar{\mu}}{\bar{\mu}}
\gamma_q(\alpha_s(\bar{\mu}))\;.
\label{16}
\end{equation}
$K_{0}$ is the modified Bessel function of order zero, which comes from the
Fourier transform to $b$ space of the gluon propagator. Note that we insert
the two-loop $\alpha_s$ in Eq.~(\ref{ral}) into the above integral of
$\gamma_q$, while it is the one-loop $\alpha_s$ that was employed in
\cite{LS}. The exponential
$e^{-S}$ decreases rapidly at large $b$, leading to Sudakov suppression
mentioned before. The essential advantage of Eq.~(\ref{15}), compared to
Eq.~(\ref{pi}), is then the extra $b$ dependence in the hard scattering.
If $b$ is small, radiative corrections with the argument of $\alpha_s$ set
to $t$ will be small, regardless of the values of $x$. Of course, when
$b$ is large and $x_1x_2Q^2$ is small, radiative corrections are still large.
However, we shall show that $e^{-S}$ suppresses the contributions from this
region, and the perturbation theory becomes relatively self-consistent.

In Eq.~(\ref{15}) the evolution of $\phi$ with $1/b$ is
written as \cite{Ji}
\begin{equation}
\phi(x,1/b)=\frac{3f_\pi}{\sqrt{2N_c}} x(1-x)
\left[\sum_{n=0}^{\infty}a_nC^{3/2}_n(2x-1)
\left(\frac{\alpha_s(1/b)}{\alpha_s(\mu_0)}\right)^{\gamma_n}
\right]\;,
\label{phie1}
\end{equation}
where the coefficients $a_n$ and the exponents $\gamma_n$ are
\begin{eqnarray}
a_n&=&\frac{5(2n+3)}{(n+2)(n+1)}I_n\;,
\\
\gamma_n&=&\frac{4}{33-2n_f}\left[1+4\sum_{k=2}^{n+1}\frac{1}{k}-
\frac{2}{(n+1)(n+2)}\right]\;,
\end{eqnarray}
with $I_0=2/15$, $I_1=0$, $I_2=8/35$ and $I_n=0$ or $n\ge 3$. The relevant
Gagenbauer polynomials are $C^{3/2}_0(x)=1$ and $C^{3/2}_2(x)=
(15/2)(x^2-1/5)$. Then Eq.~(\ref{phie1}) reduces to
\begin{equation}
\phi(x,1/b)=\frac{3}{\sqrt{2N_c}}f_\pi x(1-x)
\left[1+[5(1-2x)^2-1]\left(\frac{\alpha_s({\tilde \mu})}
{\alpha_s(\mu_0)}\right)^{\frac{50}{81}}\right]\;,
\label{phie}
\end{equation}
where we have changed the argument of $\alpha_s$ from $1/b$ to
\begin{equation}
{\tilde \mu}=\max(1/b,m)\;,
\end{equation}
with $m$ a mass scale greater than the initial value $\mu_0=0.5$ GeV. This
modification, implying that $\phi(x,1/b)$ is ``frozen" at $\phi(x,m)$
as $1/b < m$, guarantees Eq.~(\ref{phie}) to evolve toward the correct
direction. Without this freezing scale, $\phi(x,1/b)$ is ill-defined for
$1/b\to \Lambda$, and large infrared contributions to the pion form factor
will be observed. $m$ will be treated as a parameter of order $\mu_0$, and
determined by the data fitting. Obviously, Eq.~(\ref{phie}) approaches the
asymptotic wave function $\phi^{AS}$ in Eq.~(\ref{as}) as $1/b \to \infty$,
and the preasymptotic wave function $\phi^{CZ}$ in Eq.~(\ref{cz}) as
$1/b\to \mu_0$. We then expect that the pion form factor derived from
Eq.~(\ref{phie}) is located at between those from $\phi^{AS}$ and
$\phi^{CZ}$.

We are also interested in the full expression of $H$ in Eq.~({\ref{hna}),
except for the approximate one in Eq.~(\ref{7}). Without neglecting the
virtual quark transverse momemta, the parallel formula to Eq.~(\ref{15})
involves a double-$b$ integral:
\begin{eqnarray}
F_{\pi}(Q^2)&=&16\pi{\cal C}_F \int_{0}^{1}d x_{1}d x_{2}
\int_{0}^{\infty} b_{1}d b_{1}b_{2}d b_{2}\phi(x_{1},1/b_1)\phi(x_{2},1/b_2)
\nonumber \\
&\times&\alpha_{s}(t)K_{0}(\sqrt{x_{1}x_{2}}Qb_{1})
\,\exp\left[-S(x_{1},x_{2},b_{1},b_{2},Q)\right]
\nonumber \\
&\times& \left[\theta(b_{1}-b_{2})
K_{0}(\sqrt{x_{1}}Qb_{1})I_{0}(\sqrt{x_{1}}Qb_{2})\right.+
\nonumber \\
& &\left.\theta(b_{2}-b_{1})K_{0}(\sqrt{x_{1}}Qb_{2})I_{0}(\sqrt{x_{1}}Qb_{1})
\right]\; ,
\label{f5}
\end{eqnarray}
with $I_0$ the modified Bessel function of order zero, and the complete
Sudakov exponent
\begin{eqnarray}
S&=&\sum_{i=1}^2 s(x_{i},b_{i},Q)
+s(1-x_{i},b_{i},Q)+2\int_{1/b_i}^{t}\frac{d\bar{\mu}}{\bar{\mu}}
\gamma_q(\alpha_s(\bar{\mu}))\;,
\end{eqnarray}
and the hard scale
\begin{equation}
t={\rm max}\left[\sqrt{x_1x_2}Q, 1/b_1, 1/b_2\right]\; .
\end{equation}
A freezing scale is not necessary for Eq.~(\ref{f5}), since the
extra degrees of freedom $b_2$ moderate the infrared enhancement from the
low end of the evolution of the wave functions. This will be explicitly 
demonstrated in next Section.
\vskip 1.0cm

\centerline{\large \bf III. Numerical Results}
\vskip 0.5cm

In the numerical analysis the exponential $e^{-S}$ will be treated as
follows. In the small $b$ region higher-order corrections should be absorbed
into the hard amplitude, instead of into the wave functions, giving their
evolution in $b$. Hence, we set any Sudakov factor
$e^{\textstyle -s(\xi,b,Q)}$ to unity for $\xi Q/\sqrt{2} < 1/b$. In this
vein, we also set $e^{-S}$ to unity, whenever it exceeds unity in the small
$b$ region. As $b$ increases, $e^{-S}$ decreases, reaching zero at
$b=1/\Lambda$. Suppression in the large $b$ region is weaker for smaller $Q$,
where Sudakov effects are mild. The above typical behavior of $e^{-S}$ has
been shown in \cite{LS}. It is expected that the suppression improves the
self-consistency of the perturbative calculation of the pion form factor.

We evaluate $Q^2F_\pi(Q^2)$ employing the pion wave function $\phi$ with
evolution in Eq.~(\ref{phie}). We find, at $Q^2=9$ GeV$^2$,
$Q^2F_\pi(Q^2)=0.33$ GeV$^2$ for the freezing scale $m=\mu_0$ and
$Q^2F_\pi(Q^2)=0.28$ GeV$^2$ for $m=2\mu_0$, the latter being only 15\%
smaller than the former. It implies that our formalism is insensitive to
the choice of $m$. If a freezing scale was not introduced, the predictions
are almost enlarged by a factor 6, which is attributed to the infrared
enhancement from the region with $1/b < \mu_0$, {\it i.e.}, to the misuse of
Eq.~(\ref{phie}). We adopt the natural choice $m=\mu_0$ in the analysis
below.

As to the self-consistency of the perturbative calculation, we propose to
analyze Eq.~(\ref{15}), not by cutting off the $x$ integrals near their
endpoints \cite{R}, but by testing the sensitivity of the integral to the
large $b$ region. We expect that, because Eq.~(\ref{pi}) is the true
asymptotic behavior, as $Q$ increases, the $b$ integral will become more and
more dominated by the small $b$ contributions, for which the effective
coupling is small.
To see how the contributions to Eq.~(\ref{15}) are distributed in $b$ space
under Sudakov suppression, the integration is done with a variable cut off
in $b$, $b_{c}$. Typical numerical results are displayed in Fig.~3 for the
use of $\phi$ in Eq.~(\ref{phie}) with $\Lambda=0.15$ GeV. The curves,
showing the dependence of $Q^{2}F_{\pi}(Q^2)$ on $b_{c}$,
rise from zero at $b_c=0$, and reach their full height at $b_{c}=1/\Lambda$,
beyond which we consider any remaining contributions as being truly
nonperturbative. As anticipated, we observe a faster rise as $Q$ increases.

To be quantitive, we may consider the cutoff up to which half of the
whole contribution has been accumulated. For $Q^2=4$ and 9 GeV$^2$, 50\% of
$Q^{2}F_{\pi}(Q^2)$ comes from the regions with $b \leq 4.0$ GeV$^{-1}$
[$\alpha_s(1/b)/\pi \leq 0.44$] and with $b \leq 3.2$ GeV$^{-1}$
[$\alpha_s(1/b)/\pi \leq 0.37$], respectively.
A liberal standard to judge the relevance of the perturbative method
would be that 50\% of the results come from the region where
$\alpha_s(1/b)/\pi$ is no larger than, say, 0.5. From this point of view,
PQCD begins to be self-consistent at $Q^2\sim 4$ GeV$^2$. This
conclusion is consistent with that drawn in \cite{LS}, where the CZ wave
function $\phi^{CZ}$ and the asymptotic wave function $\phi^{AS}$ were
employed.

Results of $Q^2F_\pi(Q^2)$ for the use of $\phi^{AS}$, $\phi^{CZ}$ and
$\phi$ in Eq.~(\ref{phie}), as well as the experimental data
\cite{d1,d2,d3,CJB}, are shown in Fig.~4. As expected, the curve
corresponding to Eq.~(\ref{phie}) is located at between the other two.
It is known that the pion form factor behaves like $1/Q^2$ asymptotically.
The data points of $Q^2F_\pi(Q^2)$, arising from low $Q^2$ and reaching a
pleatau at $Q^2 \sim 2$ GeV$^2$, reflect this asymptotic behavior.
Hence, the point at $Q^2\sim 10$ GeV$^2$, jumping to a value about
twice of the pleatau, seems not plausible, and is not considered here.
In this sense, our predictions are in a good agreement with the data
for $Q^2 > 4$ GeV$^2$.

For $Q^2 < 4$ GeV$^2$, all the curves rise rapidly and deviate from the data,
indicating the dominance of the nonperturbative contrubutions. Note that
the predictions from $\phi^{AS}$ and $\phi^{CZ}$ are larger than those
obtained in \cite{LS}. This is due to the insertion of the two-loop
$\alpha_s$ into the integral of $\gamma_q$ in Eq.~(\ref{16}). To examine
the stability of our predictions to the variation of the scales $\Lambda$
and $\mu_0$, we compute $Q^2F_\pi(Q^2)$ for $\Lambda=0.1$ and 0.2 GeV
($\mu_0=0.5$ GeV), and for $\mu_0=0.4$ and 0.6 GeV ($\Lambda=0.15$ GeV) at 
$Q^2=9$ GeV$^2$. Their values are listed in
Table I. Apparently, the difference from those corresponding to
$\Lambda=0.15$ GeV and $\mu_0=0.5$ GeV is small.

At last, we evaluate $Q^2F_\pi(Q^2)$ using the double-$b$ factorization
formula in Eq.~(\ref{f5}}), and the results are presented in Fig.~5. Those
derived from the single-$b$ formula in Eq.~(\ref{15}) and the data are also
displayed for comparision. We find that the two curves are close to each
other especially in the region with large $Q^2\sim 5$-10 GeV$^2$. The
consistency of the predictions from Eq.~(\ref{f5}) with the data is
nontrivial, since it does not involve an adjustable freezing scale $m$. The
above analyses indicate that our scheme of truncating the infrared
enhancement from the evolution of the wave function indeed makes sense.
In the future study of the proton form factor, we shall take the same
approach in order to simply the analysis, that is, neglect the transverse
momenta carried by the virtual quarks, and introduce a freezing scale for
the proton wave function.

\vskip 1.0cm

\centerline{\large \bf IV. Conclusion}
\vskip 0.5cm

In this paper we have performed a very thorough investigation of the pion
form factor based on the modified PQCD formalism. Both the two approaches
with the single-$b$ formula plus the introduction of a freezing scale and
with the double-$b$ formula lead to results that match the experimental
data. We have also confirmed that our predictions are dominated by
perturbative contributions for $Q^2 > 4$ GeV$^2$, and insensitive to the
variation of the parameters $\Lambda$ and $\mu_0$. All these conclusions
imply that our formalism is reliable for the study of exclusive QCD
reactions at intermediate energy scales. We shall extend it to more
complicated processes, such as the proton form factor and Compton scattering,
which will be published elsewhere.

\newpage
Table I. The values of $Q^2F_\pi(Q^2)$ for different $\Lambda$ and $\mu_0$
at $Q^2=9$ GeV$^2$.
\[ \begin{array}{cccc}
\hline\hline
\Lambda({\rm GeV}) & Q^2F_\pi(Q^2)({\rm GeV}^2) &
\mu_0({\rm GeV}) & Q^2F_\pi(Q^2)({\rm GeV}^2)  \\
\hline
0.10        & 0.2974        & 0.4  & 0.3114         \\
0.15      & 0.3309        & 0.5    & 0.3309      \\
0.20      & 0.3659        & 0.6    & 0.3333      \\
\hline\hline
\end{array}    \]

\newpage
\centerline{\large \bf Figure Captions}
\vskip 0.5cm

\noindent
{\bf Fig. 1.} (a) Factorization of the pion form factor. The symbol
$\times$ represents the photon vertex. (b) Basic scattering diagrams.
\vskip 0.5cm

\noindent
{\bf Fig. 2.} Radiative corrections to the basic scattering diagrams.
\vskip 0.5cm

\noindent
{\bf Fig. 3.} Dependence of $Q^{2}F_{\pi}(Q^{2})$ on $b_{c}$ for
the use of $\phi$ in Eq.~(\ref{phie}) at $Q^2=4$ GeV$^2$ (dashed
line) and at $Q^2=9$ GeV$^2$ (solid line).
\vskip 0.5cm

\noindent
{\bf Fig. 4.} Dependence of $Q^{2}F_{\pi}(Q^{2})$ on $Q^2$ for the use
of $\phi^{AS}$ (dotted line), of $\phi^{CZ}$ (dashed line), and of $\phi$ in
Eq.~(\ref{phie}) (solid line).
\vskip 0.5cm

\noindent
{\bf Fig. 5.} Dependence of $Q^{2}F_{\pi}(Q^{2})$ on $Q^2$ for the use
of $\phi$ in Eq.~(\ref{phie}) derived from Eq.~(\ref{15})
(dashed line) and from Eq.~(\ref{f5}) (solid line).

\end{document}